\documentclass[aps, prb, superscriptaddress, twocolumn, showpacs, longbibliography, footnoography]{revtex4-2}
\usepackage{graphicx}
\usepackage{amsmath,amssymb}
\usepackage{mathtools,mathrsfs,dsfont}
\usepackage{bm, physics}
\usepackage[dvipsnames]{xcolor}
\usepackage[colorlinks, linkcolor=Red, citecolor=NavyBlue, urlcolor=NavyBlue]{hyperref}
\usepackage[all]{hypcap} % let hyperlinks point to figs not captions
\usepackage{enumitem}% http://ctan.org/pkg/enumitem
\usepackage{verbatim}
\usepackage{soul} %for strikeout text
\usepackage{array, makecell}
\usepackage{calc}% http://ctan.org/pkg/calc
\usepackage{blkarray}
\usepackage{tikz}
\usetikzlibrary{positioning}

\definecolor{cool_green}{rgb}{0.0, 0.5, 0.0}
\definecolor{new_pink}{rgb}{1, 0.0, 1}
\definecolor{blue}{rgb}{0, 0, 1}

\newcommand{\onesmatrix}{%
  \begin{array}{cccc}
    1 & \cdots & 1 \\
    \vdots & \ddots & \vdots \\
    1 & \cdots & 1
  \end{array}%
}

\begin{document}

\title{Higher-form entanglement asymmetry and topological order}

\author{Amanda Gatto Lamas}\email{amandag6@illinois.edu}
\author{Jacopo Gliozzi}\email{jglioz2@illinois.edu}
\author{Taylor L. Hughes}
\affiliation{Department of Physics and Anthony J. Leggett Institute for Condensed Matter Theory, University of Illinois at Urbana-Champaign, Urbana,  IL 61801, USA}
\date{\today}
	
\begin{abstract}
We extend a recently defined measure of symmetry breaking, the entanglement asymmetry, to higher-form symmetries. In particular, we focus on Abelian topological order in two dimensions, which spontaneously breaks a 1-form symmetry. Using the toric code as a primary example, we compute the entanglement asymmetry and compare it to the topological entanglement entropy. We find that while the two quantities are not strictly equivalent, both are sub-leading corrections to the area law and can serve as order parameters for the topological phase. We generalize our results to Abelian topological order and express the maximal entanglement asymmetry in terms of the {number of anyon species}. Finally, we discuss how the scaling of entanglement asymmetry correctly detects topological order in the deformed toric code, where 1-form symmetry breaking persists even in a non-topological phase.
\end{abstract}

\maketitle

\section{Introduction}The Landau paradigm classifies phases of matter through local order parameters and how they transform under global  symmetries. In the last decade, this framework has been extended to include higher-form symmetries, whose generators act on lower-dimensional submanifolds of space~\cite{kapustin_higherform_2014, gaiotto_higherform_2015}. Ordinary (0-form) symmetry operators act on all of space at once and measure point charges, while higher-form symmetries measure extended objects, such as string charges. As a result, higher-form symmetries have been widely used to study systems that have nonlocal excitations, like gauge theories~\cite{kapustin_chapter_2017, gaiotto_theta_2017, kitano_theta_2017, lake_ssb_2018, hsin_comments_2019, hofman_photon_2019, gomes_introduction_2023, mcgreevy_review_2023}.

Many properties of conventional symmetries also extend to higher-form symmetries. For example, both types of symmetries can be spontaneously broken. Spontaneous symmetry-breaking (SSB) of 0-form symmetries leads to conventionally ordered phases such as ferromagnets and superfluids, which can be detected locally via order parameters and correlation functions. Topological order was thought to lie outside the SSB paradigm, but Abelian topological order has been reinterpreted as SSB of a higher-form symmetry~\cite{nussinov_symmetry_2009, nussinov_symmetry2_2009, wen_emergent_2019, mcgreevy_review_2023,xu_gauge_2025}. In this context, anyon worldlines act as symmetry operators that can transform between topologically degenerate ground states. As topological order features nonlocal correlations, it is natural to wonder if this higher-form symmetry breaking can be detected locally. 

To answer this question, we employ a recently developed measure of symmetry breaking, the \textit{entanglement asymmetry}~\cite{ares_asymmetry_2023}. Defined as a relative entropy between a density matrix and its symmetrized version, the entanglement asymmetry quantifies how much the breaking of a global symmetry can be detected in a subregion. 
Initially developed to measure symmetry restoration after a quantum quench~\cite{ares_asymmetry_2023,rylands_mpemba_2024, joshi_experiment_2024}, the asymmetry has now been calculated for a variety of quantum states and (0-form) symmetry groups~\cite{capizzi_ising_2023, ares_u1_2023, capizzi_mps_2024, murciano_xy_2024,yamashika_freefermion_2024, fossati_defects_2024,caceffo_dissipative_2024, liu_circuit_2024, fujimura_nonabelian_2025,Benini_2025}. 

In this work, we generalize the entanglement asymmetry to higher-form symmetries. We develop a method to compute the asymmetry for Abelian topological order and use the toric code~\cite{kitaev_anyons_2006}, whose ground states spontaneously break a 1-form symmetry, as an explicit example. Moreover, we link this asymmetry to the topological entanglement entropy (TEE), which detects the long-range entanglement patterns of topological order~\cite{hamma_tee_2005, levin_tee_2006, kitaev_tee_2006}. Both quantities originate from the anyon structure of the phase, and we clarify their subtle differences.
Finally, we compute the asymmetry of the deformed toric code~\cite{castelnovo_quantum_2008}, which has 1-form SSB that persists even in a non-topological phase~\cite{huxford_gaining_2023}. We demonstrate that the  entanglement asymmetry in the non-topological phase vanishes in the thermodynamic limit, and thus find that the {scaling of the} entanglement asymmetry may be used as a probe of topological order in the presence of 1-form SSB.

\begin{figure}[t!]
\label{fig:subregion}
\includegraphics[width=\linewidth]{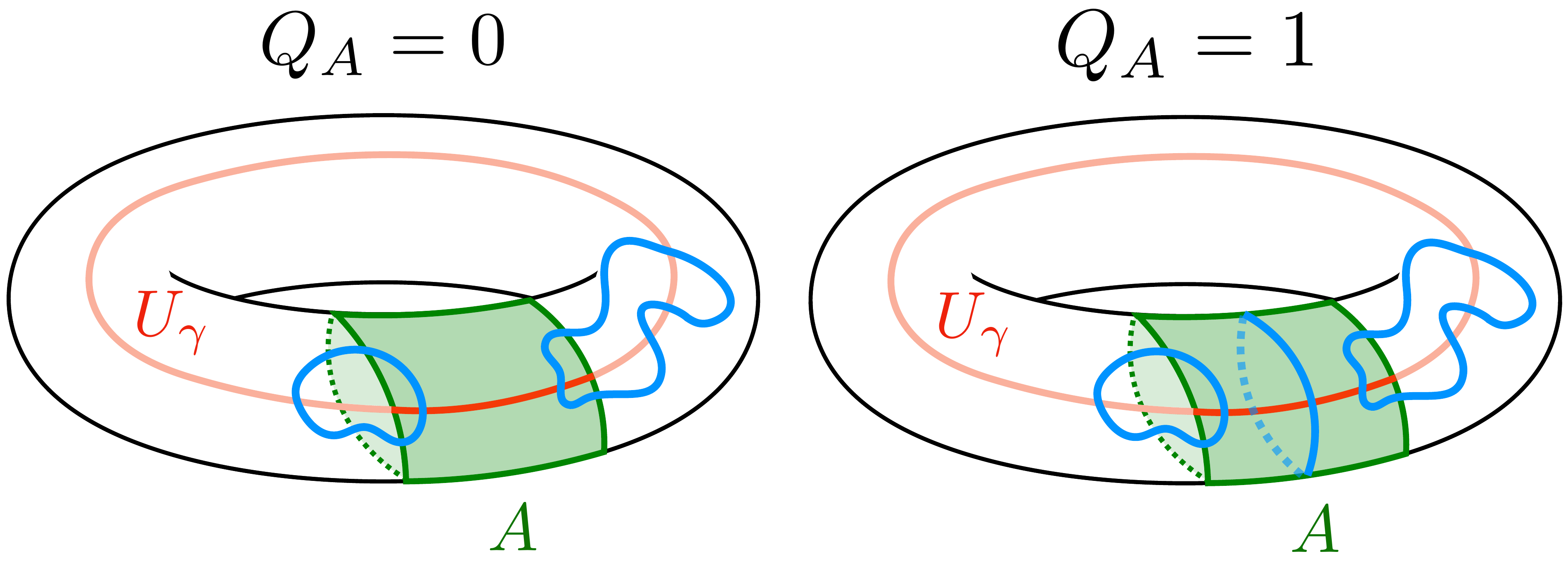}
    \caption{A 1-form symmetry operator (red) and its string charges (blue) on a torus. Inside a cylindrical subregion $A$ (green), the total charge measured by the symmetry operator can take different values. The dark (light) red part of the symmetry operator represents its factorization into region A(B). %\jg{Hopefully this ugly space goes away when comments are removed}
    }
\end{figure}

\section{Higher-form symmetry}Symmetries in many-body quantum systems act via unitary operators. Specifically, a $p$-form symmetry in $D$ spatial dimensions is generated by unitary operators on a closed $(D - p)$-dimensional manifold~\cite{kapustin_higherform_2014, gaiotto_higherform_2015}. The charges of a $p$-form symmetry are $p$-dimensional, e.g., point charges for $p=0$ and strings for $p=1$. In what follows, we focus on 1-form symmetries in $D=2$ spatial dimensions.

To illustrate, consider a 0-form U$(1)$ symmetry. Associated with the symmetry is a 1-form current $j$ that satisfies the conservation law $d \star j = 0$, where $\star$ is the Hodge star and $d$ is the exterior derivative. Integrating over a closed spatial manifold $\mathcal{M}$ yields the total charge, $Q_\mathcal{M} = \int_{\mathcal{M}} \star j$, which counts the number of point charges in $\mathcal{M}$. The symmetry operator is $U_\mathcal{M}(\alpha) = e^{i \alpha Q_\mathcal{M}}$, where $\alpha \in [0, 2\pi)$ is a phase in U$(1)$. The conservation law implies that the manifold $\mathcal{M}$ can be smoothly deformed without changing $U_\mathcal{M}$, so the symmetry operator is said to be topological.

For a 1-form U$(1)$ symmetry, the current is a conserved 2-form. The total charge, $Q_\gamma = \int_{\gamma} \star j$, is then defined on a closed one-dimensional manifold $\gamma$. Here the operator $Q_\gamma$ counts the number of one-dimensional string charges that intersect $\gamma$.
Again, the associated symmetry operator $U_\gamma(\alpha) = e^{i \alpha Q_\gamma}$ is topological. As a result, the number of string charges intersecting $\gamma$ does not change if $\gamma$ is deformed. An illustration of 1-form symmetry operators and their charges is shown in Fig.~\ref{fig:subregion}.

Here we instead focus on discrete 1-form symmetries where the symmetry operators are similarly topological, despite not having exactly conserved currents~\cite{gaiotto_higherform_2015}.   
One setting in which discrete 1-form symmetries arise is Abelian topological order, where closed anyon lines act as symmetry operators~\cite{wen_emergent_2019, mcgreevy_review_2023,xu_gauge_2025}. Indeed, these operators form a group determined by the fusion of anyons, and can be smoothly deformed in spacetime. On topologically nontrivial spatial manifolds, systems with topological order have degenerate ground states that differ by the insertion of non-contractible anyon lines. In the symmetry picture, this is 1-form SSB, since each ground state transforms to another under the action of the symmetry.

\section{Entanglement asymmetry}Our goal is to quantify the degree of higher-form symmetry breaking by generalizing the entanglement asymmetry~\cite{ares_asymmetry_2023}. To gain intuition, let us first review the definition of entanglement asymmetry for 0-form symmetries.
In the simplest scenario, entanglement asymmetry probes how much a state $\ket{\psi}$ breaks a discrete 0-form symmetry $G$ in a spatial subregion $A$. We assume that the unitary operators representing this symmetry factorize over space as $U_\mathcal{M}(g) = U_A(g) \otimes U_{B}(g)$, where $B$ is the spatial complement of $A$. The state on $A$ is given by the reduced density matrix $\rho_A = \text{Tr}_{B}(\ket{\psi}\bra{\psi})$. This state is said to be symmetric if $U_A(g) \rho_A U_A(g)^\dag = \rho_A$ for all $g\in G$~\footnote{More precisely, the mixed state $\rho_A$ is weakly symmetric under the restriction of $G$ to the subregion $A$.}. 

To measure symmetry-breaking, we compare $\rho_A$ to its symmetrized version:
\begin{equation}\label{eq:rho_sym}
    \rho_A^\text{sym} = \frac{1}{|G|} \sum_{g\in G} U_A (g) \rho_A U_A(g)^\dag,
\end{equation}
where $|G|$ is the {order} of the group. While $\rho_A$ may mix symmetry sectors with different charges in $A$, $\rho_A^\text{sym}$ is block diagonal in the symmetry basis.
\begin{figure}[t!]
\label{fig:exceptions}
\includegraphics[width=\linewidth]{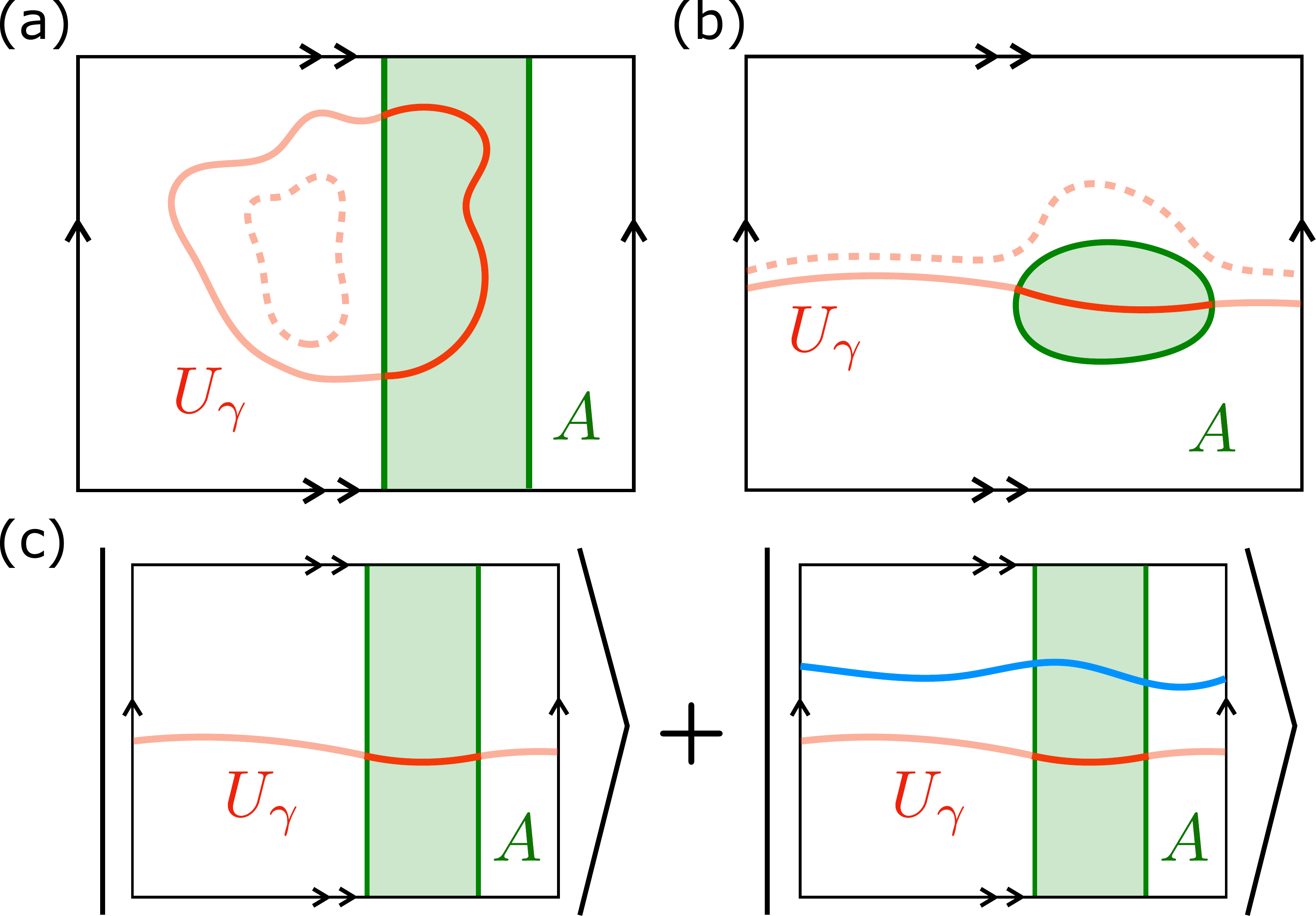}
\caption{A 1-form symmetry on a torus with trivial asymmetry, $\Delta S_A = 0$. (a) Contractible symmetry operator $U_\gamma$, (b) Contractible subregion $A$, (c) Symmetry-breaking in the horizontal direction (superposition of different numbers of blue charges) cannot be detected by a symmetry operator in the same direction, such as $U_\gamma$.
}
\end{figure}
The entanglement asymmetry is then defined as
\begin{equation}\label{eq:def_ent_asym}
    \Delta S_A = S[\rho_A^\text{sym}]-S[\rho_A],
\end{equation}
where $S[\rho] = - \text{Tr}(\rho \log \rho)$ is the von Neumann entanglement entropy.
Heuristically, $\Delta S_A$ captures the information contained about global symmetry-breaking in subregion $A$.
The asymmetry is always nonnegative, as Eq.~\eqref{eq:def_ent_asym} is the relative entropy between $\rho_A$ and $\rho_A^\text{sym}$~\cite{ares_asymmetry_2023}. Moreover, $\Delta S_A = 0$ if and only if $\rho_A$ is symmetric. 

We now extend this to the case of a higher-form symmetry. While all 0-form symmetry operators are supported on the full space $\mathcal{M}$, their higher-form analogs have support on only lower dimensional submanifolds $\gamma \subset \mathcal{M}$. Once again, we require that these operators factorize,
\begin{equation}\label{eq:factorization}
    U_\gamma(g) = U_{\gamma,A}(g) \otimes U_{\gamma, {B}}(g),
\end{equation}
where $U_{\gamma, A}$ acts only on subregion $A$. An example of this decomposition on a torus is shown in Fig.~\ref{fig:subregion}, where the dark (light) part of the red line is $U_{\gamma_x,A}$ ($U_{\gamma_x,B}$). 
The higher-form entanglement asymmetry is given by Eq.~\eqref{eq:def_ent_asym}, though with an additional dependence on the choice of support $\gamma$. 

Because the 1-form symmetry operators are topological, $\Delta S_A$ cannot change under deformations of $\gamma$. This fact imposes stringent constraints on the entanglement asymmetry. For example, a contractible $\gamma$ gives $\Delta S_A=0$, as $U_\gamma$ can be deformed to lie entirely outside any subregion $A$. Similarly, any contractible $A$ has $\Delta S_A=0$, since all $U_\gamma$ can be deformed to lie outside of $A$. These two situations are depicted in Fig.~\ref{fig:exceptions}(a-b).
Even when both $A$ and $\gamma$ are non-contractible, a state may break a higher-form symmetry along one manifold $\gamma'$, while preserving the same symmetry along a topologically distinct $\gamma$. For example, the state depicted in Fig.~\ref{fig:exceptions}(c) is a superposition of different string charges in the horizontal direction, but these charges cannot be detected by a symmetry operator along the same direction.
Each of these cases of exactly zero asymmetry are a direct result of the lower-dimensional support of higher-form symmetries. In what follows, we therefore take non-contractible $A$ and $\gamma$, and compute the asymmetry of states that transform under $U_\gamma$.

\section{Toric code and 1-form SSB}The toric code is defined by the Hamiltonian~\cite{kitaev_anyons_2006}
\begin{equation}\label{eq:ham_tc}
H = -\sum_{v} \, \begin{tikzpicture}[scale = 0.5, baseline = {([yshift=-.5ex]current bounding box.center)}]
    \draw[black] (1.5,0) -- (-1.5, 0);
    \draw[black] (0,1.5) -- (0,-1.5);
    \node at (0.9, 0) {\normalsize $Z$};
    \node at (-0.9, 0) {\normalsize $Z$};
    \node at (0, 0.9) {\normalsize $Z$};
    \node at (0, -0.9) {\normalsize $Z$};
    \node at (0.2, 0.2) {\small $\textcolor{black}{v}$};
\end{tikzpicture} \,
- \sum_p 
\begin{tikzpicture}[scale = 0.5, baseline = {([yshift=-.5ex]current bounding box.center)}]
        \draw[color = black] (-1, -1) -- (-1, 1) -- (1, 1) -- (1, -1) -- cycle;
        \node at (-1, 0) {\small $X$};
        \node at (1, 0) {\small $X$};
        \node at (0, -1) {\small $X$};
        \node at (0, 1) {\small $X$};
        \node at (0, 0) {\small $\textcolor{black}{p}$};
    \end{tikzpicture},
\end{equation}
where $Z$ and $X$ are Pauli operators on the links of a square lattice, and we take the spatial manifold $\mathcal{M}$ to be a torus of linear size $L$.
The Hamiltonian in Eq.~\eqref{eq:ham_tc} commutes with two types of string operators, and therefore has two 1-form symmetries. We focus on the symmetry generated by the (electric) Wilson line,
\begin{equation}\label{eq:sym}
U_\gamma = \prod_{l \in \gamma} X_l,
\end{equation}
where $\gamma$ is a closed loop on the lattice. There are two topologically distinct classes of non-contractible loops on the torus, $\gamma_x$ and $\gamma_y.$ Since the symmetry operators are topological, we can, without loss of generality, choose an element of each class to define independent symmetry operators, $U_{x}\equiv U_{\gamma_x}$ and $U_{y}\equiv U_{\gamma_y}$. Because $(U_\gamma)^2 = 1$ for any $\gamma$, the symmetry group is $\mathbb{Z}_2$.

The charges of this {1-form} symmetry are loops of $X=-1$ on the dual lattice, i.e., magnetic flux lines.
A charge on the dual loop $\tilde{\gamma}$ is created/annihilated by the charged operator
\begin{equation}\label{eq:charged_op}
    T_{\tilde{\gamma}} = \prod_{l \perp \tilde{\gamma}} Z_l.
\end{equation}
This is the (magnetic) 't Hooft line, which flips the values of $X$ on links along $\tilde{\gamma}$. In analogy with the non-contractible Wilson loops, we define $T_x\equiv T_{\tilde{\gamma}_x}$ and $T_y\equiv T_{\tilde{\gamma}_y}$.

The action of the 1-form symmetry is encoded in the relation
\begin{equation}\label{eq:sym_relation}
    U_\gamma T_{\tilde{\gamma}} U_\gamma^\dag = (-1)^{I(\gamma, \tilde{\gamma})} T_{\tilde{\gamma}}, 
\end{equation}
where $I(\gamma, \tilde{\gamma})$ is the intersection number of $\gamma$ and $\tilde{\gamma}$. The symmetry operator $U_\gamma$ acts nontrivially only on charged operators $T_{\tilde{\gamma}}$ that intersect it an odd number of times. 
It follows that $U_\gamma$ counts the number of charges intersecting $\gamma$ modulo 2 (see Fig.~\ref{fig:subregion}). We can therefore write $U_\gamma = e^{i \pi Q_\gamma}$, where $Q_\gamma$ is the total 1-form charge piercing $\gamma$. Note that symmetry operators act trivially on contractible charges, as contractible loops always cross the symmetry operators an even number of times. 

The 't Hooft lines $T_{\tilde{\gamma}}$ also commute with the Hamiltonian, and therefore are themselves symmetry operators for another 1-form symmetry. The symmetry group is again $\mathbb{Z}_2$, as $(T_{\tilde{\gamma}})^2=1$. The charges of this symmetry are loops of $Z=-1$ on the lattice, i.e., electric flux lines. These charges are created/annihilated by the Wilson line $U_\gamma$.
The full 1-form symmetry group for the toric code is then $\mathbb{Z}_2\times\mathbb{Z}_2$, and reflects the conservation of electric and magnetic flux modulo 2.

Although the symmetry group is Abelian, the symmetry operators obey the group law only up to a phase in Eq.~\eqref{eq:sym_relation}. The Wilson and 't Hooft lines therefore form a projective representation of the symmetry, which is an indication of an anomaly~\cite{gaiotto_higherform_2015, hsin_comments_2019, gomes_introduction_2023}. 
The 1-form anomaly is crucial for topological order: the projective phases between symmetry operators encode the nontrivial braiding statistics of anyons, i.e., the $e$ and $m$ anyons of the toric code~\cite{mcgreevy_review_2023, huxford_gaining_2023}.

On the torus, the ground state subspace of Eq.~\eqref{eq:ham_tc} is fourfold degenerate. In the $Z$-basis, each ground state is an equal weight superposition of all contractible loops of $Z = -1$ on the links of the lattice, i.e., electric fluxes. The ground states differ only in their non-contractible loops, which can be present or absent along the two cycles of the torus. For convenience, we label these four ground states $\ket{\Psi^Z_{ab}}$, where the binary index $a$($b$) indicates the absence or presence of a non-contractible loop along the $\hat{x}$($\hat{y}$)-direction.

These degenerate ground states in the $Z$-basis spontaneously break the 1-form symmetry generated by Eq.~\eqref{eq:sym}. In other words, $\ket{\Psi^Z_{ab}}$ transforms nontrivially under the action of the symmetry. For example, $U_{x}$ inserts a noncontractible flux loop: $U_{x} \ket{\Psi^Z_{00}} = \ket{\Psi^Z_{10}}$. As a result, we can interpret the topological ground state degeneracy of the toric code as 1-form SSB~\footnote{Our definition of SSB is state-dependent, and requires a ground state that transforms under the symmetry. This is a stronger condition than a perimeter law for the charged 't Hooft loop, $\expval{T_{\tilde{\gamma}}} \sim e^{-\text{Perim}(\tilde{\gamma})}$, which holds for any superposition of ground states.}. Since the ground states transform only under non-contractible symmetry operators, this definition of SSB requires a spatial manifold that is not simply connected.

We will now calculate the entanglement asymmetry for a toric code ground state. For a 0-form symmetry, specifying a state and a subregion $A$ uniquely determines the asymmetry.
In the 1-form case, however, the asymmetry also depends on the support of the symmetry operators. As shown in Fig.~\ref{fig:exceptions}, contractible subregions $A$ or contractible symmetry supports $\gamma$ both result in $\Delta S_A=0$. Then, for the entanglement asymmetry to capture the 1-form symmetry breaking, we will consider symmetry operators along the non-contractible loop $\gamma_x$, and a cylindrical subregion $A$ that wraps around the $\hat{y}$-direction of the torus (see Fig.~\ref{fig:subregion}). \footnote{We could also have considered a symmetry operator along $\gamma_y$ and a region $A$ wrapping around $x$-direction of the torus.}

{As an example}, we focus on the state $\ket{\Psi^Z_{00}}$.
It is convenient to rewrite this state in the $X$-basis, where it becomes an equal-weight superposition of all possible \emph{magnetic} flux loops on the dual lattice, contractible and non-contractible. That is,
\begin{equation}\label{eq:ssb_state_x}
    \ket{\Psi^Z_{00}}=\frac{1}{2}\left(\ket{\Psi^X_{00}} +\ket{\Psi^X_{10}} + \ket{\Psi^X_{01}} + \ket{\Psi^X_{11}}\right),
\end{equation}
where $\ket{\Psi^X_{ab}}$ labels the state with $a$($b$) non-contractible loops of $X=-1$ along the $\hat{x}$($\hat{y}$)-cycle of the dual lattice. These states are eigenstates of the symmetry operator:
\begin{equation}\label{eq:sym_eig}
    U_{x}\ket{\Psi^X_{ab}} = (-1)^b\ket{\Psi^X_{ab}}.
\end{equation}
In this basis, the SSB state $\ket{\Psi^{Z}_{00}}$ is an equal superposition of states with $Q_{x} = 0$ and $Q_{x} = 1$. Indeed, the first two states in Eq.~\eqref{eq:ssb_state_x} contain an even number of non-contractible charges piercing $\gamma_x$, while the last two contain an odd number.

To calculate the entanglement asymmetry, we compute the reduced density matrix $\rho_A = \Tr_B (\ket{\Psi^Z_{00}} \bra{\Psi^Z_{00}})$ .  From Eq.~\eqref{eq:ssb_state_x}, we can write our state as an equal-weight superposition of all configurations of closed magnetic flux loops, $|\Psi^Z_{00}\rangle \propto \sum_{\mathcal{C}}\ket{\mathcal{C}}$, where $\ket{\mathcal{C}}$ denotes a state with loop configuration $\mathcal{C}$. The full density matrix is a sum of
outer products of states having different closed string configurations: \begin{equation}\label{eq:rho_full}
\ket{\Psi^Z_{00}} \bra{\Psi^Z_{00}} \propto \sum_{\mathcal{C}, \mathcal{C}' \text{ closed strings}} \ket{\mathcal{C}} \bra{\mathcal{C}'}.
\end{equation}
Any loop configuration can be decomposed into parts that lie in $A$ and parts in $B$: $\ket{\mathcal{C}} = \ket{\mathcal{C}_A} \otimes \ket{\mathcal{C}_B}$. After tracing out $B$, we are left with strings in $A$:
\begin{equation}\label{eq:rhoA_toric}
    \rho_A \propto \sum_{\mathcal{C}_A, \mathcal{C}_A'}\ket{\mathcal{C}_A} \bra{\mathcal{C}'_A}.
\end{equation}
Here each configuration $\mathcal{C}_A$ can contain both closed strings and open strings that end on the boundary $\partial A.$ The latter arise since some closed loops on the full torus are split between $A$ and $B$. 

\begin{figure}[t!]
\includegraphics[width=\linewidth]{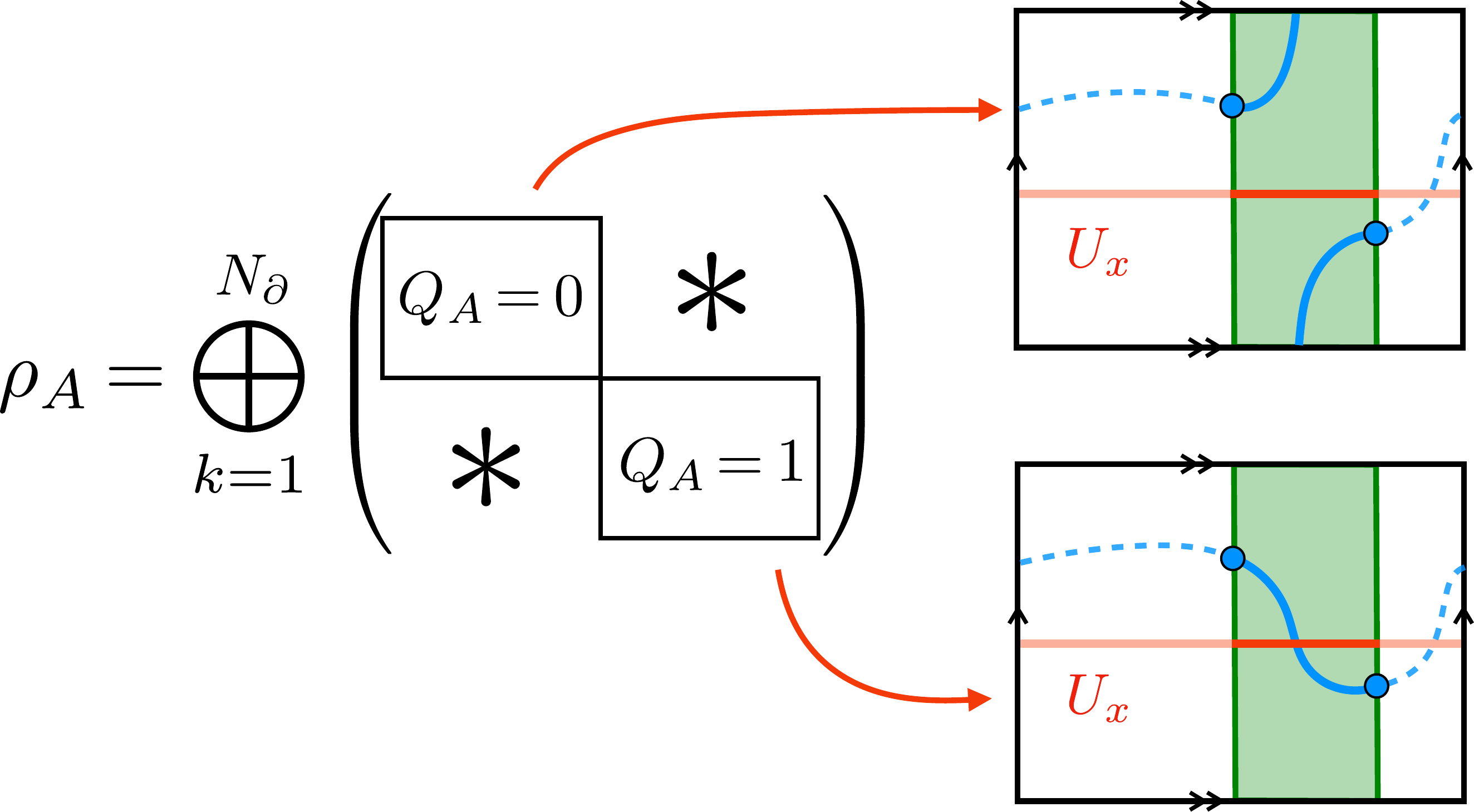}
    \caption{In a toric code ground state, $\rho_A$ splits into sectors labeled by the pattern of string charge (blue curves) intersections at the boundary of $A$. Each sector splits into diagonal blocks of fixed charge in $A$, and off-diagonal blocks that mix charges in $A$. The structure of $\rho_A$ is insensitive to deformations of the symmetry operator (red).} 
    \label{fig:sectors}
\end{figure}

The partial trace ensures that only pairs of loop configurations $\mathcal{C}$ and $\mathcal{C'}$ that are \emph{identical} in $B$ contribute to $\rho_A$. Since the boundary is shared between $A$ and $B$, every term of Eq.~\eqref{eq:rhoA_toric} is of the form $\ket{\mathcal{C}_A}\bra{\mathcal{C}'_A}$, where $\mathcal{C}_A$ and $\mathcal{C}'_A$ have the same pattern of boundary intersections.
Thus, the reduced density matrix $\rho_A$ decomposes into sectors labeled by the pattern of magnetic flux strings crossing $\partial A$~\cite{hamma_tee_2005, pretko_u1_2016}. An example of two configurations that lie in the same sector is shown in  Fig.~\ref{fig:sectors}.

In the basis of magnetic flux strings, the state $\ket{\Psi^Z_{00}}$ is an equal weight superposition of all loops. Working in this basis restricted to $A$ we find
\begin{equation}
\label{eq:rhoA_toric_block}
\rho_A \propto \bigoplus_{k = 1}^{N_\partial} \begin{pmatrix} \onesmatrix \end{pmatrix},
\end{equation}
where $k$ indexes the $N_\partial$ sectors with distinct boundary string intersections, and the size of each sector is determined by the (large) number of string configurations compatible with the boundary-intersection constraint. Each element in a sector of $\rho_A$ has a fixed pattern of boundary intersections but is distinguished by its string configuration inside of $A$. 
We can further organize each sector using the 1-form symmetry charge, noting that for every configuration with $Q_A = 0$, there is another one with $Q_A = 1$ and the same pattern of boundary intersections. 
Two such configurations drawn from the same sector are shown in Fig.~\ref{fig:sectors}, and are related by inserting a non-contractible 't Hooft loop transverse to the symmetry operator~\footnote{We note that if we deform the symmetry operator the $Q_A$ charge sector labels could flip.}.
Since the SSB state {$\ket{\Psi^Z_{00}}$} is a superposition of all loops, each block of $\rho_A$ contains an equal number of configurations with $Q_A = 0$ and $Q_A = 1$~\footnote{This can be seen from $\ket{\Psi^Z_{00}} \propto (1 + T_{\tilde{\gamma}_y})(1 + T_{\tilde{\gamma}_x})\ket{\Psi^X_{00}}$, where $T_{\tilde{\gamma}}$ is a non-contractible 't Hooft loop.}. 

The decomposition of $\rho_A$ into sectors allows us to easily compute the entanglement entropy $S[\rho_A]$. Each sector contributes a rank one matrix and thus only one nonzero eigenvalue, so the total entanglement entropy is $\log N_\partial.$
We can determine the number of sectors $N_\partial$ as follows. There are $2L$ links on the boundary of $A$, each of which can either be crossed by a string or not, giving $2^{2L}$ distinct boundary intersections. However, strings in $A$ always form closed loops in the full system, so the number of boundary crossings must be even. This constraint therefore gives $N_\partial=2^{2L-1}$ independent sectors, and hence the entanglement entropy decomposes as:
\begin{equation}
    \label{eq:EE_toric}
    S[\rho_A] = 2L \log 2 - \log 2.
\end{equation}
The first term is an area law term, which is proportional to $|\partial A| = 2L$, while second term is the topological entanglement entropy~\cite{hamma_tee_2005, levin_tee_2006, kitaev_tee_2006}.

The last step in calculating the asymmetry is to symmetrize $\rho_A$. {For our full symmetry group $\mathbb{Z}_2\times \mathbb{Z}_2$,
\begin{align}
    \rho_A^{\text{sym}}=\frac{1}{4}(\rho_A+U_{x,A}\rho_AU^\dagger_{x,A}+T_{x,A}\rho_AT^\dagger_{x,A}\nonumber\\+U_{x,A}T_{x,A}\rho_AT^\dagger_{x,A}U_{x,A}^\dagger), \label{eq: full sym rhoA}
\end{align}
where $U_{x, A}$ and $T_{x, A}$ are the symmetry operators restricted to region $A$. Since the state $|\Psi^Z_{00}\rangle$ is an eigenstate of $T_x$, $\rho_A$ is left invariant under the action of $T_{x,A}$. As a result, $\rho_A^\text{sym}$ can be written as
\begin{equation}
    \label{eq:rhoA_sym_toric}
    \rho_A^\text{sym} = \frac{1}{2} \left( \rho_A + U_{x, A}\, \rho_A \, U^{\dag}_{x, A}  \right).
\end{equation}
Here the symmetrized reduced density matrix reflects the fact that the original state $|\Psi^Z_{00}\rangle$ breaks only the $\mathbb{Z}_2$ symmetry generated by the $U_x$ operator inside region $A$, instead of the full $\mathbb{Z}_2\times \mathbb{Z}_2$ symmetry group.}

In practice, $\rho_A^\text{sym}$ may be constructed by deleting off-diagonal elements of $\rho_A$ that couple sectors having different $Q_A$ (see asterisks in Fig. \ref{fig:sectors})\cite{ares_asymmetry_2023}. 
By setting these elements to zero, we obtain $\rho_A^\text{sym}$. Since each boundary sector splits into two blocks corresponding to the symmetry sectors, the resulting entanglement entropy is $S[\rho_A^\text{sym}] = \log 2N_\partial=2L\log 2$~\footnote{Before we had $N_\partial$ blocks of  matrices with identical unit entries where each had one nonzero eigenvalue, thus $\log N_\partial$. Now we have $2N_\partial$ blocks of matrices having identical unit entries, hence each has one nonzero eigenvalue, thus $\log 2N_\partial$.}. Inserting this result into Eq.~\eqref{eq:def_ent_asym}, we obtain
\begin{equation}\label{eq: first example}
\Delta S_A = \log 2.
\end{equation}
This is the maximal asymmetry possible for the group $\mathbb{Z}_2$, indicating that one bit of information is encoded in the symmetry-breaking inside $A$. Moreover, this result matches the 0-form symmetry case, where a broken discrete symmetry $G$ results in $\Delta S_A \approx \log |G|$~\cite{capizzi_ising_2023,capizzi_mps_2024}.

Let us comment on the state dependence of this result. As in ordinary SSB, one can always {choose a symmetric basis for the ground state subspace}. Here, the symmetric states $\ket{\Psi^X_{ab}}$ all have $\Delta S_A = 0$. However, even states that break the 1-form symmetry can have zero asymmetry. For instance, the state $\frac{1}{\sqrt{2}}(\ket{\Psi^X_{00}} + \ket{\Psi^X_{10}})$, shown schematically in Fig.~\ref{fig:exceptions}(c), transforms under $U_{y}$ {and $T_y$}, but this symmetry-breaking cannot be detected by the operators $U_{x}$ {and $T_x$}.
This example highlights the need to choose a consistent subregion and symmetry operator to detect 1-form asymmetry. 

\section{Topological entanglement entropy}In the toric code example above, $S[\rho_A^\text{sym}]$ encodes the area-law part of $S[\rho_A]$, while $\Delta S_A=\log 2$ extracts a subleading correction. This latter term is none other than the topological entanglement entropy (TEE), $S_\text{topo}$~\cite{hamma_tee_2005, levin_tee_2006, kitaev_tee_2006}. While the area law coefficient is model-dependent, the TEE encodes the long-range entanglement pattern of the topological phase. It is then natural to ask if the 1-form entanglement asymmetry and the TEE are the same. 

While there are connections between the two quantities that we now illustrate, the asymmetry and TEE are different in general. Conceptually, the TEE reflects the fact that all flux strings in the ground state of the toric code must form closed loops. This constraint reduces the entanglement entropy of any subregion, contractible or non-contractible. In contrast, the entanglement asymmetry is a measure of \textit{symmetry breaking}, and captures the presence of different 1-form charges, i.e., non-contractible strings, inside a non-contractible subregion. 

We can highlight the differences between the TEE and the entanglement asymmetry from some simple  examples. In a contractible region of the toric code, $S_\text{topo}$ is bounded below by $\log 2$~\cite{kim_bound_2023}, but $\Delta S_A$ vanishes. If the boundary of region $A$ is decorated with a 1D SPT phase, $S_\text{topo}$ can increase {spuriously} ~\cite{cano_cut_2015, zou_spurious_2016, williamson_spurious_2019, kim_bound_2023}, but $\Delta S_A$ remains unchanged, as it depends only on non-contractible loops inside $A$. 
{Finally, for disconnected subregions, the TEE grows with the number of boundaries~\cite{zhang_quasiparticle_2012}, while the asymmetry is bounded by the order of the 1-form symmetry group.}

Despite their differences, the 1-form entanglement asymmetry and TEE can be {directly} related {for toric code ground states on a non-contractible cylinder subregion.}
To see this, we first focus on a the special class of so-called \textit{minimum entropy states} (MESs)~\cite{zhang_quasiparticle_2012, dong_tee_2008}. The MESs are ground states of the toric code that {minimize the entanglement entropy for a non-contractible region $A$ by maximizing the negative correction to the area law in that region -- that is, by maximizing $S_\text{topo}$.} In these states, $S_\text{topo}=n_{\partial A}\log 2$, where $n_{\partial A}$ is the number of boundaries of region $A$. {We will now see that the MESs also maximize the entanglement asymmetry.}

The MESs are defined as simultaneous eigenstates of the Wilson and 't Hooft loops {which wrap around the non-contractible cycle inside subregion $A$. For now on, we refer to these Wilson and 't Hooft loop operators  
as running ``along the entanglement cut'', as in Ref.~\cite{zhang_quasiparticle_2012}.} 
The 't Hooft loop along the cut for our cylindrical region $A$ in Fig.~\ref{fig:subregion} is $T_y$. An eigenstate of $T_y$ must be a linear combination of the basis states $|\Psi^X_{ab}\rangle$ having the same number of non-contractible loops in the $x$-direction, and different non-contractible loops in the $y$-direction. That is, the state must have the form
\begin{equation}\label{eq: MES cond 1}
    |\text{MES}_y\rangle = \frac{1}{\sqrt{2}}\left(\ket{\Psi^X_{ab}}\pm T_y\ket{\Psi^X_{ab}}\right),
\end{equation}
{which is also an eigenstate of $U_y$.}
Note that these states break the symmetry in the correct orientation for $\Delta S_A$ to capture the symmetry breaking (cf. superpositions such as in Fig.~\ref{fig:exceptions}(c) which break symmetry in the incorrect orientation). {Since the MESs form a basis for the ground space subspace, any state may be written as a linear combination of the MESs, and the TEE for a non-contractible entanglement cut} 
{explicitly depends}
{on this linear combination~\cite{zhang_quasiparticle_2012}. Next, we will show that the 1-form entanglement asymmetry exactly captures this TEE for any ground state of the toric code.}

{First, consider a generic state which is a linear superposition of the different MESs, and note that different MESs have orthogonal support in the complement of the non-contractible region $A$ ~\cite{zhang_quasiparticle_2012}. Then the partial trace over the complement of $A$ yields a reduced density matrix $\rho_A$ which is a convex combination of the form 
\begin{equation}
    \rho_A=\sum_{i=1}^4 c_i\rho_A^{\text{MES}_y^i}, \label{eq: convex comb MES}
\end{equation}
  where $\rho_A^{\text{MES}_y^i}$ are the reduced density matrices of the $i^{\text{th}}$ MES with respect to region $A$, and $c_i\geq0$ are coefficients obeying $\sum_i c_i=1$.} 
  {The entanglement entropy of $\rho_A$ is then~\cite{Nielsen_Chuang_2010}
\begin{align}
    S(\rho_A)&=\sum_ic_i[S(\rho_A^{\text{MES}_y^i})-\log c_i],\\
    &=S(\rho_A^{\text{MES}})+H(\{c_i\}),
\end{align}
where $H(\{c_i\})=-\sum_ic_i\log c_i\geq 0$
is the Shannon entropy, and we used the fact that $S(\rho^\text{MES}_A)=S(\rho_A^{\text{MES}_y^i})$ is the same for all MESs.}

{The next step to calculate the asymmetry is to obtain $\rho_A^\text{sym}$ as in Eq. \ref{eq: full sym rhoA}, reproduced here for convenience:}
\begin{align*}
    \rho_A^{\text{sym}}=\frac{1}{4}(\rho_A+U_{x,A}\rho_AU^\dagger_{x,A}+T_{x,A}\rho_AT^\dagger_{x,A}\\ +U_{x,A}T_{x,A}\rho_AT^\dagger_{x,A}U_{x,A}^\dagger). 
\end{align*}
{The key observation is that, since the MESs are simultaneous eigenstates of $U_y$ and $T_y$, they transform under both $T_x$ and $U_x$. In fact, $T_x$ and $U_x$ permute the different states $|\text{MES}_y\rangle$, as can be seen from Eq. \eqref{eq: MES cond 1}. This contrasts with the previous example, where $|\Psi^Z_{00}\rangle$ transformed under only $U_x$.}
{As a result, the MESs spontaneously break both $\mathbb{Z}_2$ symmetries of the toric code: the magnetic one generated by $U_x$ and the electric one generated by $T_x$.}

{Using the fact that the symmetry operators permute the MESs and expanding in the $\{\rho_A^{\text{MES}_y^i}\}$ basis,}
{we can write $\rho_A^\text{sym}$ as the equal superposition
\begin{align}
    \rho_A^\text{sym} = \frac{1}{4}\sum_i\rho_A^{\text{MES}_y^i}.
\end{align}
Then, $\rho_A^\text{sym}$ has the \textit{maximal} entanglement entropy  
\begin{equation}
    S(\rho_A^\text{sym})=S(\rho_A^\text{MES})+2\log 2.
\end{equation}
Finally, the 1-form asymmetry is given by 
\begin{equation}
    \Delta S_A(\rho_A) =2\log 2-H(\{c_i\}) .\label{eq: asym general}
\end{equation}
Note that this is the same result obtained for the TEE in Ref. \cite{zhang_quasiparticle_2012}.} 
{Here, we have provided a symmetry-based interpretation of this formula for the correction to the entanglement entropy of a non-contractible region.}

{Indeed, the entanglement asymmetry makes the role of 1-form SSB explicit in characterizing topological order through entanglement}. {Other works, in particular Ref. \cite{xu_gauge_2025}, have also studied how the 1-form SSB contributes nontrivially to the TEE. In Ref. \cite{xu_gauge_2025}, the authors compare the TEE of ground states which spontaneously break a 1-form symmetry to the TEE of those which do not. While that work shares the spirit of our analysis, the procedure that we present to calculate $\Delta S_A$ is different than the comparison made in Ref. \cite{xu_gauge_2025}.} 
{Our work is then complementary to Refs. \cite{zhang_quasiparticle_2012} and \cite{xu_gauge_2025}, and also serves to highlight the conceptual differences between the TEE and measures of symmetry breaking. Indeed, although the 1-form asymmetry takes the same values as the TEE when region $A$ has only one connected non-contractible component, the asymmetry and the TEE are conceptually different quantities, and may take different values when other types of regions are considered, as previously discussed.}

{To gain intuition about Eq.~\eqref{eq: asym general}, we now comment on a few special cases and relate this formula back to our previous examples. First, note that the entanglement asymmetry in Eq.~\eqref{eq: asym general} is maximized when $c_i=1$ for some $i$, which kills the $H(\{c_i\})$ contribution. Taking $c_i=1$ corresponds to considering the reduced density matrix of an MES, and we obtain the maximum $\Delta S(\rho_A^\text{MES})=2\log 2$. Just as in the previous example for $|\Psi_{00}^Z\rangle$, this result matches the expectation that $\Delta S_A(\rho_A^\text{MES})=\log |G|$, where $G=\mathbb{Z}_2\times \mathbb{Z}_2$ is the total 1-form symmetry group which is \textit{maximally} spontaneously broken by the MESs, while in our previous example we had $G=\mathbb{Z}_2$ for the state $|\Psi^Z_{00}\rangle$.}

{Importantly, the MESs in Eq.~\eqref{eq: MES cond 1}, which are defined with respect to a non-contractible subregion $A = A_y$ wrapping around the $y$-cycle of the torus (see Fig.~\ref{fig:subregion}), maximize only the asymmetry for this subregion $A_y$. To see this, consider their asymmetry in the subregion $A_x$, a cylinder wrapping around the $x$-cycle of the torus.
By construction, the MESs in Eq.~\eqref{eq: MES cond 1} are symmetric under $U_y$ and $T_y$, the two symmetry operators that intersect $A_x$, i.e., that cannot be deformed outside of $A_x$. As a result, their entanglement asymmetry in this region, $\Delta S_{A_x}$, is exactly zero. The MESs therefore maximally break the symmetry in $A_y$, but do not break it at all in $A_x$.}

{The relationship between asymmetries in different directions is also evident in our earlier example, $|\Psi_{00}^Z\rangle$ in Eq.~\eqref{eq:ssb_state_x}. This state breaks only one of the two $\mathbb{Z}_2$ symmetries, but it does so in \emph{both} directions of the torus. Specifically, $|\Psi_{00}^Z\rangle$ transforms under both $U_x$ and $U_y$, but not under $T_x$ and $T_y$. As a result, the entanglement asymmetry for this state is the same for both non-contractible cylinders: $\Delta S_{A_x} = \Delta S_{A_y} = \log 2$. On the other hand, MESs have an asymmetry of $2\log 2$ in one direction and zero along the other. In both cases, the sum of asymmetries in opposite directions is bounded by $2 \log 2$.}

{Using the connection between asymmetry and TEE in Eq.~\eqref{eq: asym general}, the bound of $2 \log 2$ extends to any ground state of the toric code. In Ref.~\onlinecite{zhang_quasiparticle_2012}, this bound was likened to an ``uncertainty principle'' for the TEE, as it constrains how much information about a state can be extracted in different non-contractible subregions of the torus. Entanglement asymmetry further sharpens this analogy. First, it gives a physical meaning to the information contained in different subregions, which is simply the amount of symmetry-breaking. Second, it connects the uncertainty principle to a non-commutation relation between operators. As we saw with the MES, to maximally break the symmetry generated by $U_x$ and $T_x$, a state must be symmetric under $U_y$ and $T_y$. As a result, the asymmmetry cannot be maximized in both directions at once. Fundamentally, this bound on asymmetry and TEE follows from the fact that symmetry operators in different directions, like $U_x$ and $T_y$, do not commute, {{and therefore do not share eigenstates}}.~\footnote{The anticommutation of Abelian symmetry operators is a sign that the 1-form symmetry acts projectively, and is therefore anomalous.}.}

{To conclude our discussion of asymmetry and TEE, we note that our result for the toric code can be extended to any Abelian topological order in two spatial dimensions. Here, the relevant symmetry is an Abelian group $G$ generated by all possible anyon lines.}
{For concreteness, we consider the cylindrical subregion  $A_y$, as in Fig.~\ref{fig:subregion}.
Once again, the MESs are defined as the simultaneous eigenstates of all non-contractible anyon lines along the entanglement cut ~\cite{zhang_quasiparticle_2012}. 
These lines are simply the (commuting) 1-form symmetry operators along the $y$-cycle of the torus $U_y$ and $T_y$.}

{We are interested in the asymmetry of the MESs under symmetry operators that cannot be deformed out of subregion $A_y$. 
These operators are the anyon lines along the $x$-cycle of the torus, which are transverse to $A_y$.
Such anyon lines intersect only an odd number of times with the anyon lines along the $y$-cycle,
and can therefore act on them nontrivially. More precisely, braiding non-degeneracy implies that for every transverse anyon line $a_x$, there exists an anyon line $b_y$ along the entanglement cut such that $[a_x, b_y] \neq 0$~\cite{kitaev_anyons_2006}. Since the MESs are eigenstates of all $b_y$ along the cut, they must transform under the full symmetry group generated by the symmetry operators $a_x$. In other words, the MESs maximally break the symmetry $G$.} 
{We can then repeat our procedure to obtain an equation analogous to Eq.~\eqref{eq: asym general} for $\Delta S_A(\rho_A)$. Again, the 1-form entanglement asymmetry is maximized for the MES, with $\Delta S_A^\text{max}=\log|G|$}
{counting the number of anyon species in the topological order.}

\section{Deformed toric code} Thus far, we have treated 1-form SSB interchangeably with topological order.  However, Ref.~\cite{huxford_gaining_2023} recently challenged this intuition, using the deformed toric code~\cite{castelnovo_quantum_2008} as {an example of 1-form SSB without topological order.}  
This model is defined by adding a term to the Hamiltonian in Eq.~\eqref{eq:ham_tc} of the form
\begin{equation}\label{eq:ham_deformed}
    H_d=\sum_v e^{-\beta \sum_{l \ni v}X_l},
\end{equation}
where the exponentiated sum is over the four links {(the star)} that contain the vertex $v$. Crucially, the deformation $H_d$ still commutes with the 1-form symmetry operators $U_\gamma${, while $T_{\bar{\gamma}}$ is no longer an exact symmetry for $\beta>0$}.
The deformed toric code has a topological phase with $S_\text{topo}=\log 2$ for $\beta<\beta_c$, { and a non-topological phase with $S_\text{topo}=0$ for $\beta>\beta_c$ \cite{huxford_gaining_2023}.}
Remarkably, both phases have the same ground state degeneracy as the $\mathbb{Z}_2$ toric code, i.e., four ground states on a torus.

In Ref.~\cite{huxford_gaining_2023}, this degeneracy was argued to be a result of 1-form SSB in both phases. Indeed, the 1-form symmetry is spontaneously broken for any $\beta$, and one might expect $\Delta S_A>0$ for both the topological and non-topological phases of the deformed toric code. If this were the case, the entanglement asymmetry would \textit{not} be an accurate probe for topological order. 
However, while it is true that $\Delta S_A>0$ for any $\beta$ at finite system size, we now argue that the scaling of $\Delta S_A$  in the thermodynamic limit distinguishes the topological and non-topological phases {of the deformed toric code}. Concretely, for a fixed-width, non-contractible region A, $\Delta S_A$ vanishes as $L\rightarrow \infty$ in the non-topological phase, while it remains nonzero in the topological phase.

To demonstrate this result, let us consider the ground states of the deformed toric code. For any $\beta$, the ground states of the deformed toric code $|\Psi^X_{ab}(\beta)\rangle$ can be obtained from the undeformed toric code ground states $|\Psi^X_{ab}(0)\rangle$ as
\begin{equation}
    |\Psi^X_{ab}(\beta)\rangle = \frac{1}{\sqrt{N_{ab}(\beta)}}S(\beta)|\Psi^X_{ab}(0)\rangle,
\end{equation}
where $S(\beta)=\prod_l e^{\beta X_l/2}$, and $N_{ab}(\beta)$ is a normalization factor \cite{castelnovo_quantum_2008,huxford_gaining_2023}. Note that the operator $S(\beta)$ applies a decreasing weight to string configurations in $|\Psi^X_{ab}(0)\rangle$ that have longer $X=-1$ (magnetic flux) loops. 

We will focus on the entanglement asymmetry of the state 
\begin{equation}\label{eq:deformed_mes}
    |\text{MES}_+(\beta)\rangle=\frac{1}{\sqrt{2}}\left(|\Psi^X_{00}(\beta)\rangle +|\Psi^X_{01}(\beta)\rangle \right)
\end{equation}
for different values of $\beta.$ First, we consider the non-topological phase as $\beta \rightarrow \infty$.
In this limit, $|\Psi^X_{00}(\beta \rightarrow \infty) \rangle = |+ + \cdots \rangle$ is the product state of $X=+1$ on every link. The other ground states, which have non-contractible flux loops, are forced by $H_d$ to be superpositions of non-contractible loops that are only {straight lines}. For instance, 
\begin{equation}\label{eq:psi_01_beta}
    |\Psi^X_{01}(\beta \rightarrow \infty)\rangle=\frac{1}{\sqrt{L}}\sum_{i}^{L}T_{y}^{(i)}|+ + \cdots \rangle,
\end{equation}
where $T_{y}^{(i)}$ inserts a non-contractible flux loop parallel to the $\hat{y}$-direction at position $x=i$. As $L\to\infty$, a vanishing proportion of states in the superposition of Eq.~\eqref{eq:psi_01_beta}, and thus also in $\ket{\text{MES}_+(\beta\rightarrow \infty)}$, host a non-contractible loop inside the region $A$. 
Consequently, {the proportion of states with $Q_A=1$ that appear in $\rho_A$ is also vanishing}. The reduced density matrix then acquires a definite charge $Q_A=0$ in the thermodynamic limit, and is therefore symmetric, giving $\Delta S_A = 0$.

For large but finite $\beta>\beta_c$, states with small contractible loops have nonzero weights in $\ket{\text{MES}_+(\beta)}$. However, adding contractible loops inside of $A$ does not change $Q_A$. The only way for two states to have different values of $Q_A$ is for them to differ by a non-contractible loop of length $L$ inside region $A$. As in the $\beta\rightarrow \infty$ case, these non-contractible loops are increasingly rare in the thermodynamic limit, so the reduced density matrix $\rho_A$ remains approximately symmetric, with $\Delta S_A \approx 0$.

Next, we consider the topological phase at $\beta < \beta_c$.
For small but nonzero $\beta$, the ground states are similar to those of the ordinary toric code. In particular, $\ket{\text{MES}_+(\beta)}$ is a superposition of all magnetic flux loops, with coefficients that slightly penalize longer loops. In a typical loop configuration, 
$\mathcal{O}(L^2)$ of the links are covered by a flux line. Inserting a non-contractible loop in $A$ flips only an $\mathcal{O}(L)$ number of links, leaving $\mathcal{O}(L^2)$ occupied links. It follows that states with $Q_A=0$ and $Q_A=1$ have roughly the same number of occupied links, and therefore roughly the same coefficient in $\ket{\text{MES}_+(\beta)}$. The entanglement asymmetry of this state is then $\Delta S_A \approx \log 2$.  {Additionally, in Appendix \ref{app:numerics}, we provide numerical evidence that the asymmetry vanishes in the non-topological phase, and approaches $\log 2$ in the topological phase as the thermodynamic limit is taken.} We thus find that the entanglement asymmetry can distinguish the phases of the deformed toric code in the thermodynamic limit.

In Ref.~\cite{huxford_gaining_2023}, the authors argue that the distinguishing feature between the topological and the non-topological phase of the deformed toric code is the presence of an \textit{anomalous} 1-form symmetry in the topological phase, which is absent in the non-topological phase. Recall that in the undeformed toric code, both Wilson loops $U_\gamma$ and 't Hooft loops $T_{\tilde{\gamma}}$ are 1-form symmetry operators. Their anticommutation is then a sign of a mixed anomaly. As discussed above, the Wilson loop $U_\gamma$ is still a 1-form symmetry of the deformed toric code, but the fate of the 1-form symmetry generated by the 't Hooft loop depends on $\beta.$
For $\beta < \beta_c$, a deformed version of the 't Hooft loop remains a symmetry of the ground state subspace and anticommutes with $U_\gamma,$ thus preserving the anomaly.
However, when $\beta > \beta_c$, this deformed 't Hooft loop becomes non-unitary, and is therefore no longer a symmetry.

Without a second 1-form symmetry in addition to the Wilson loop, there is no mixed anomaly in the {non-topological} phase. Intriguingly, our results suggest that the scaling of $\Delta S_A$ in the thermodynamic limit is sensitive to this subtle anomaly structure. We leave the precise connection between the 1-form anomaly and the ability of the entanglement asymmetry to capture topological order to future work. {Here, we merely comment that the absence of the 't Hooft loop symmetry in the non-topological phase is the common source of both the absence of the anomaly and the vanishing of $\Delta S_A$ in the thermodynamic limit.}

\section{Conclusion}
By generalizing the entanglement asymmetry, we have quantified 1-form symmetry breaking in topologically ordered phases. In particular, we calculated the asymmetry for ground states of the toric code and related it to the topological entanglement entropy. Furthermore, we argued that this entanglement asymmetry serves as a faithful probe of topological order. Indeed, the asymmetry tends to zero in the non-topological phase of the deformed toric code, where 1-form SSB occurs without topological order.

{An important limitation of our work is that the calculation of the asymmetry requires knowledge of how the symmetry operators act on the state, which is not always clear away from the fixed point of the topological order, where the 1-form symmetry may be emergent in the ground state. While the deformed toric code is an example of a model with an emergent 1-form symmetry for $0<\beta<\beta_{c}$, it is a particularly well-characterized one \cite{castelnovo_quantum_2008,huxford_gaining_2023}, which allows for determining the asymmetry. It would be interesting to consider other practical ways of calculating the entanglement asymmetry for models generally away from the fixed point.}

Another natural extension to our work
% that we have not mentioned 
is defining the entanglement asymmetry for non-invertible symmetries, which emerge in non-abelian topological order~\cite{shao_noninvertible_2023}.
It is also interesting to test our results in theories with continuous 1-form symmetry, such as U$(1)$ gauge theory without charges. In such a theory the deconfined phase in three dimensions spontaneously breaks a U$(1)$ 1-form symmetry, with the photon as the Goldstone boson~\cite{gaiotto_higherform_2015}.
Based on the number of symmetry sectors, the entanglement asymmetry should behave as $\Delta S_A \sim \log L$, which intriguingly matches the subleading topological correction to the entanglement entropy in U(1) gauge theory~\cite{pretko_u1_2016}.

Entanglement asymmetry has also proven useful to quantify how dynamics can locally restore a broken conventional symmetry. It would be interesting to consider whether higher-form symmetries, having non-local order parameters, may still be \textit{locally} restored after a quench. 
One possibility is a 1-form symmetry analog of the quantum Mpemba effect~\cite{ares_asymmetry_2023,rylands_mpemba_2024,murciano_xy_2024,joshi_experiment_2024}, whereby states that break the symmetry the most are also the fastest to restore it. 
However, as higher-form SSB can encode a quantum memory, a Mpemba effect would be undesirable for quantum information storage~\cite{dennis_memory_2002}. Instead, classes of states that maintain 1-form SSB for long times are ideal for quantum computation. Careful studies of 1-form SSB dynamics may therefore be useful for applications in topological quantum computing. 

\textit{Note added}.---During the preparation of this manuscript, we became aware of two independent works expanding the definition of the entanglement asymmetry to non-invertible and higher-form symmetries from a high energy perspective~\cite{ahmad_asymmetry_2025, benini_asymmetry_2025}.

\textit{Data availability}. ---The source code and parameters used to generate Fig. \ref{fig:numerics} are publicly available \cite{zenodo_jgliozzi}.

%%%%%%%%%%%%%%%%%%%%%%%%%%%%%
% ACKNOWLEDGEMENTS
%%%%%%%%%%%%%%%%%%%%%%%%%%%%%
\noindent
\textit{Acknowledgements}.--- A.G.L. thanks John McGreevy for a useful conversation. J.G. and T.L.H. acknowledge support from the US Office of Naval Research MURI grant N00014-20-1-2325. A.G.L. is supported by the 2025-26 AAUW International Fellowship. 
\\

\bibliography{higher_form_EA.bib}

\setcounter{figure}{0}
\renewcommand{\thefigure}{S\arabic{figure}}
\setcounter{equation}{0}
\renewcommand{\theequation}{S\arabic{equation}}

\newpage
\appendix
{
\section{Numerical calculation of entanglement asymmetry in the deformed toric code}\label{app:numerics}

We have argued in the main text that the scaling of entanglement asymmetry with system size serves as a signature of topological order in the deformed toric code. In particular, for a fixed-width, non-contractible subregion $A$, the asymmetry $\Delta S_A$ remains finite in the thermodynamic limit ($L \rightarrow \infty$) in the topological phase, while it goes to zero in the non-topological phase. Although our arguments focused on the limits $\beta \rightarrow 0$ (deep in the topological phase) and $\beta\rightarrow\infty$ (deep in the trivial phase), we now provide numerical evidence that $\Delta S_A$ is a genuine order parameter for any deformation strength $\beta$.

Consider an $L_x \times L_y$ square lattice with periodic boundary conditions in both directions, and a  cylindrical subregion $A$ of size $\ell_x \times L_y$, where the width $\ell_x$ is fixed (see inset of Fig.~\ref{fig:numerics}). We are interested in the ground states of the deformed toric code Hamiltonian in Eq.~\eqref{eq:ham_deformed} for different values of the deformation parameter $\beta$. In particular, we consider ground states like Eq.~\eqref{eq:deformed_mes}, reproduced below for convenience:
\begin{equation}\label{eq:deformed_mes_again}
    |\text{MES}_+(\beta)\rangle=\frac{1}{\sqrt{2}}\left(|\Psi^X_{00}(\beta)\rangle +|\Psi^X_{01}(\beta)\rangle \right).
\end{equation}
Such states spontaneously break the 1-form symmetry generated by $U_x$, the Wilson loop along the $x$-cycle of the torus. Specifically, $|\text{MES}_+(\beta)\rangle$ transforms to an orthogonal ground state when acted on by $U_x$, such that $\langle \text{MES}_+(\beta)|U_x|\text{MES}_+(\beta)\rangle = 0$.

\begin{figure}[b!]
\label{fig:numerics}
\includegraphics[width=0.95\linewidth]{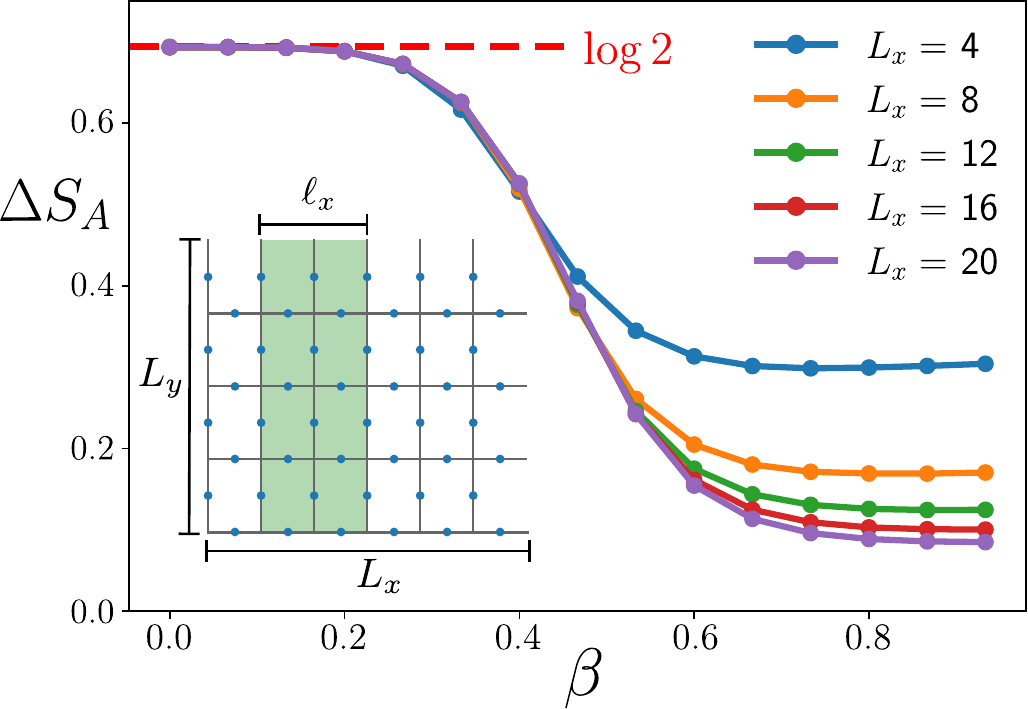}
    \caption{Entanglement asymmetry $\Delta S_A$ of the deformed toric code for a cylindrical subregion of the torus, calculated using MPS approximations of the symmetry-breaking ground states. For small values of the deformation $\beta$, the asymmetry is close to $\log 2$ for all system sizes $L_x$, while for large values of $\beta$ it decays with increasing $L_x$. Inset: Cylindrical subregion of width $\ell_x = 2$ and height $L_y=4$ shaded in green.
    }
\end{figure}

Using the density matrix renormalization group (DMRG), we obtain matrix product state (MPS) approximations of Eq.~\eqref{eq:deformed_mes_again} at different values of $\beta$. This procedure is slightly subtle because of the four degenerate ground states on the torus. To isolate the correct ground state, we explicitly insert Wilson loops along the $x$- and $y$-cycles of the torus as external fields in the deformed toric code Hamiltonian:
\begin{equation}\label{eq:ham_external_field}
H(\beta) \mapsto H(\beta) - J_x U_x - J_y U_y.
\end{equation}
Since $U_x$ and $U_y$ commute with $H(\beta)$, their addition does not change the eigenstates of the model. However, $J_x$ and $J_y$ apply an energetic shift between different topological sectors. For instance, when $J_x, J_y > 0$, the unique ground state of Eq.~\eqref{eq:ham_external_field} is $|\Psi^X_{00}(\beta)\rangle$, which has eigenvalue $+1$ under both $U_x$ and $U_y$. Setting $J_x < 0$ and $J_y > 0$ leads to $|\Psi^X_{01}(\beta)\rangle$, which, when superposed with $|\Psi^X_{00}(\beta)\rangle$, gives the SSB ground state of Eq.~\eqref{eq:deformed_mes_again}. 

We apply this procedure to obtain MPS ground states on a torus of height $L_y = 4$ and variable length $L_x = 4, 8, 12, 16, 20$. We use bond dimensions of up to $\chi = 256$, and we check that $\langle U_x \rangle = 0$ for these ground states, in accordance with Eq.~\eqref{eq:deformed_mes_again}. After obtaining the correct ground state for a given $\beta$, we contract the MPS to construct the full density matrix inside a cylindrical subregion $A$ of height $L_y$ and width $\ell_x = 2$ (green shaded region in inset of Fig.~\ref{fig:numerics}). We then compare the entanglement entropy of this density matrix, $\rho_A$, to the entanglement entropy of its symmetrized version, $\rho_A^\text{sym} = \frac{1}{2} \left(\rho_A + U_{x, A} \rho_A U_{x,A} \right)$, to extract $\Delta S_A$. Throughout, all DMRG simulations were performed using the TeNPy tensor network library~\cite{tenpy}.

The results of the numerical simulations are shown in Fig.~\ref{fig:numerics}. In the topological phase, $\beta < \beta_c \approx 0.44$, the entanglement asymmetry is close to its fixed point value of $\log 2$ for all system sizes $L_x$. On the other hand, the non-topological phase with $\beta > \beta_c$ exhibits an asymmetry that falls off with system size. Note that the observed $\beta_c\approx0.44$ in our numerical simulations matches the analytical predictions in Refs. \cite{castelnovo_quantum_2008,huxford_gaining_2023}. Although we are limited to rather small system and subsystem sizes, these results provide ulterior evidence that the phases of the deformed toric code can be distinguished by the scaling of their entanglement asymmetry.
}

\end{document}